\documentclass[journal]{IEEEtran}	
\usepackage[T1]{fontenc}
\usepackage[utf8]{inputenc}
\usepackage{graphics}
\usepackage{graphicx}	
\usepackage{amsmath, amsfonts, amssymb}
\usepackage{cancel}
\usepackage{epstopdf}
\usepackage{chngcntr}
\usepackage{enumerate}
\usepackage{amsthm}
\usepackage{hyperref}
\usepackage{caption}
\usepackage{subcaption}
\usepackage{float}
\usepackage{comment}
\usepackage{empheq}
\usepackage{multirow}
\usepackage{multicol}
\usepackage[usenames]{color}
\usepackage{pgf}
\usepackage{blindtext}
\usepackage{cite}

\graphicspath{{img/}} 

\newtheorem{theorem}{Theorem}
\newtheorem{definition}{Definition}
\newtheorem{lemma}{Lemma}
\newtheorem{prop}{Proposition}

\def\N{\mathbb{N}} % Números naturales
\def\C{\mathbb{C}} % Numeros Complejos
\def\one{\mathbf{1}}
\def\P{\mathbb{P}} % Probabilidad 

\def\matT{\mathbf{T}}
\def\matI{\mathbf{I}} 
\def\matU{\mathbf{U}}
\def\matV{\mathbf{V}}

\def\matP{\mathbf{P}}
\def\matQ{\mathbf{Q}}
\def\matE{\mathbf{E}} 
\def\matC{\mathbf{C}}

\def\matA{\mathbf{A}}
\def\matW{\mathbf{W}} 

\def\matB{\mathbf{B}}

\def\Thetab{\boldsymbol{\Theta}} 
\def\Sigmab{\boldsymbol{\Sigma}}
 
\def\Lambdab{\boldsymbol{\Lambda}}

\def\u{\mathbf{u}}
\def\v{\mathbf{v}}

\def\x{\mathbf{x}}
\def\y{\mathbf{y}}
\def\w{\mathbf{w}}

\def\H{\mathcal{H}}  
\def\setX{\mathcal{X}}
\def\setY{\mathcal{Y}}

\def\matProj{\mathcal{P}}

\def\rank{\mathrm{rank}}   
  
\begin{document}							 
	
\title{Model Order Estimation for A Sum of Complex Exponentials}

\author{Raymundo Albert, Cecilia G. Galarza}
%thanks{}
\markboth{This work has been submitted to the IEEE Transaction on Signal Processing.}{Shell \MakeLowercase{\textit{et al.}}: Bare Demo of IEEEtran.cls for IEEE Journals}
%\markboth{Journal of \LaTeX\ Class Files,~Vol.~14, No.~8, August~2015}%
%{Shell \MakeLowercase{\textit{et al.}}: Bare Demo of IEEEtran.cls for IEEE Journals}

\maketitle

\begin{abstract}
 In this paper, we present a new method for estimating the number of terms in a sum of exponentially damped sinusoids embedded in noise. In particular, we propose to combine the shift-invariance property of the Hankel matrix associated with the signal with a constraint over its singular values to penalize small order estimations. With this new methodology, the algebraic and statistical structures of the Hankel matrix are considered. The new order estimation technique shows significant improvements over subspace-based methods. In particular, when a good separation between the noise and the signal subspaces is not possible, the new methodology outperforms known techniques. We evaluate the performance of our method using numerical experiments and comparing its performance with previous results found in the literature.
\end{abstract}


\IEEEpeerreviewmaketitle

\section{Introduction}

	\IEEEPARstart{A}{} ubiquitous problem in signal processing is to recover useful information from a signal modeled as a sum of complex exponentials.  This problem is significant in applications such as radar \cite{Sarkar2000}, spectroscopy \cite{Gudmundson2012}, and music signals \cite{Laroche1993}, to mention just a few. The signal to be detected usually contains unknown parameters such as amplitude, phase, frequency, etc. Subspace-based techniques \cite{Stoica2005} have shown good performance for estimating the model parameters by solving spectral estimation problems. Moreover, recent studies based in convex optimization have reported new procedures that exhibit good performance under different setups \cite{Andersson2014,Grussler2018}. Nevertheless, a sensible step in any parametric spectral estimation method is to accurately estimate the model order.
	
	%During the past decades, there has been published abundant literature on the problem of model order estimation.
	The seminal works applied information-theoretic criteria to estimate the model order\cite{Stoica2004}. The Akaike Information Criterion, the Minimum Description Length, as well as a more recent approach developed in \cite{Mariani2015}, guarantee good performance in the asymptotic case. However, for short data records, these methods are no longer optimal and they loose performance when the Signal-to-Noise Ratio (SNR) is low. 
	
	An alternative strategy uses Kronecker's theorem that states a one-to-one correspondence between a linear combination of $r$ complex exponentials and a Hankel matrix with rank $r$. Unfortunately, this result is difficult to apply in real-life implementations,  because noise contaminates the observed signal. In consequence, the Hankel matrix has full rank. Using a Singular Value Decomposition (SVD) of the Hankel matrix, it is possible to decompose its columns space in a dominant subspace related to the signal and a secondary subspace known as the noise-subspace.  The dimension of the signal subspace is established by the number of prominent singular values.  In the low SNR regime, there is no clear cut between singular values. Then, determining which ones are the relevant singular values becomes a difficult task.   Recently, the authors in \cite{Gavish2014,Gavish2017}  have addressed this problem. In particular, they have studied a non-random matrix perturbed by a noise matrix with zero mean independent and identically distributed (i.i.d) entries. For this case, they proposed a universal threshold to separate the dominant singular values of the observed matrix. In their presentation, they analyzed the statistical behavior of the singular values of Gaussian matrices.  Another method along these lines includes a detection strategy that takes into account the statistical properties of eigenvalues of Gaussian matrices  \cite{Kritchman2009}. These approaches are attractive, and they show good performance when the Signal-to-Noise Ratio (SNR) is low. However, they have poor performance when applied to Hankel matrix because they discard the statistics induced by the Hankel structure.
	
	In \cite{Qiao2020} another hard threshold for singular values was proposed for random real matrices with subgaussians entries and Toeplitz structure. When dealing with random Hankel matrix with Gaussian entries, a similar bound can be found using concentration inequalities  \cite{tropp2015}. More general random Hankel matrices were studied in  \cite{Hokanson2020}. The authors found the hard threshold as an upper bound on the spectral norm of the random Hankel matrix. Nonetheless,  this is a conservative bound that underestimates the matrix rank.

	The methods mentioned above select the model order by analyzing the spectral properties of the additive noise. Alternatively, other methods exploit the structure of the data. In the case of a Hankel matrix contructed from a sum of complex exponentials, the rotational invariance is a well-known principle employed in spectrum estimation techniques \cite{Roy1989}. The authors in \cite{Badeau2004} followed this path to propose the order selection technique known as ESTER.  Although they showed good performance when combined with the algorithm ESPRIT for spectrum estimation, ESTER is based on noiseless assumptions. An alternative technique was proposed in \cite{Papy2007} and it was called SAMOS.  While SAMOS is more robust than ESTER, both techniques work well in the high SNR regime, but they fail when the signal is not strong enough.

    In this work, we analyze some of the pitfalls of these schemes and propose a new alternative that is resilient to high noise power, while keeping it accurate when the signal gets stronger. In particular, we propose to combine the shift-invariance property of the Hankel matrix associated with the sum of exponentials with a constraint on the singular values associated with noise in a single optimization problem. In this way, we are taking into account not only the algebraic structure of the signal but also its statistical properties.
    
   The rest of the paper is organized as follows: in section \ref{sec:ModelDescription} we introduce the signal model and present the rotational invariance property. Section \ref{sec:review} reviews some techniques for model order estimation and points out their drawbacks in the context of the model introduced before. In section \ref{sec:RandomHankel} we find bounds for the spectral norm of a random Hankel matrix. Section \ref{sec:SelectionRule} introduces our proposal. In section \ref{sec:results} we perform Montecarlo simulation to compare the performance of our proposed method with other popular. Finally section \ref{sec:Conclusions} concludes the paper with final remarks.
    
	\subsection{Notation}
	
		Throughout the paper we use standard notation: lowercase ($v$) for scalars, boldface lowercase ($\v$) for vectors, uppercase boldface $\matA$ for matrices. Given a matrix $\matA$, we denote its transpose, Hermitian, and Moore-Penrose pseudo-inverse as $\matA^T$, $\matA^H$, $\matA^\dagger$ respectively. %We also refers to $\|\matA\|_F$ to the Frobenius norm of a matrix $\matA$. 
		$\|\matA\|_2$ is reserved for the induced $2-$norm. The notation $\matI_m$ us used for the $m\times m$ identity matrix.
		We use calligraphy letters ($\setX$) for subspaces. %We denote $\mathrm{dist}(\y,\setX) = \min_{\x\in\setX}\|\y-\x\|_2$ the distance of a vector $\y$ to $\setX$. Furthermore, $\matProj_\setX$ denotes the orthogonal projection onto the subspace $\setX$.
		
\section{Model Description}\label{sec:ModelDescription}

	Consider the following model
	\begin{equation}
		y_k = x_k + w_k, \qquad k = 0,1,\ldots,
		\label{Eq:noisySignal}
	\end{equation}
	where $x_k$ is given by
	\begin{equation}
		x_k = \sum_{i=1}^{r}c_iz_i^k  \qquad k = 0,1,\ldots, 
		\label{Eq:Signal_1}
	\end{equation} 
	$w_k$ is a circularly symmetric complex Gaussian process, $w_k\sim \mathcal{CN}(0,\eta^2)$, $z_i\in\C$ is a complex resonant frequency, and $c_i\in\C$ the amplitude associated with it. The goal is to estimate $r$ using the samples $y_k$, $k = 0,1,\ldots,N-1$. An appropriate model order estimation is key for an accurate estimation of the resonances $z_i$.
	
	Given $m,n>0$, define the $m\times n$-Hankel matrix, $\H_\x$ obtained from $x_0, \ldots, x_{m+n-2}$
	 \begin{equation}
		\H_{\x} = \begin{bmatrix} x_0 & x_1 & \cdots & x_{n-1} \\[0.3em]
		x_1 & x_2 & \cdots & x_{n-2} \\[0.3em]
		\vdots &  &        & \vdots \\[0.3em]
		x_{m-1} & x_{m-2}  & \cdots & x_{m+n-2}
		\end{bmatrix}.
		\label{Eq:HankelMatrix_y}
	\end{equation}
	Since $x_k$ satisfies \eqref{Eq:Signal_1}, we know that the rank of $\H_\x$ is $r$. Now, consider the Singular Value Decomposition (SVD) 
	\begin{equation}
	    \H_{\x} = \matU\Lambdab\matV, 
	    \label{Eq:SVDHx}
	\end{equation}
	where $\Lambdab$ is a diagonal matrix that contains the singular values arranged in decreasing order. Then, $\matU(r)$, which contains the first $r$ singular vectors of $\H_\x$, spans the signal space. Let us define the following matrices
	\begin{equation}
	    \begin{aligned} 
		& \matU_f(r) = \begin{bmatrix} \mathbf{0}_{(m-1)\times 1} & \matI_{m-1}
		\end{bmatrix}\matU(r), \\[0.3em]
		& \matU_l(r) = \begin{bmatrix}  \matI_{m-1} & \mathbf{0}_{(m-1)\times 1}
		\end{bmatrix}\matU(r).
		\end{aligned}
		\label{Eq:matrices_lf}
	\end{equation}
	Efficient spectral estimation techniques such as ESPRIT  \cite{Razavilar1998}, exploit the rotational invariance property of  $\H_{\x}$,
	\begin{equation}
			%\matU_f\v_i = z_i\matU_l\v_i,\quad i =1, \ldots, r
			\matU_f(r) = \matU_l(r)\boldsymbol{\Phi}.
			\label{Um:eq}
	\end{equation}
Here $\boldsymbol{\Phi} \in \C^{r \times r}$ is a non-singular matrix. According to \eqref{Um:eq}, $\matU_f(r)$ and $\matU_l(r)$ span the same subspace. Then, if $\H_\x$ were available, a plausible order estimation approach would be to find the integer $r$ that satisfies \eqref{Um:eq}. Unfortunately, the signal $x_k$ is observed only through a noisy version  $y_k$ as in \eqref{Eq:noisySignal}. Let $\H_\y$ be the 
$m\times n$-Hankel matrix built from $y_0, \ldots, y_{m+n-2}$.  
\begin{equation}
\H_{\y} = \H_{\x} + \H_{\w},
\label{Eq:HyHxHw}
\end{equation} 
where $\H_{\w}$ is a perturbation matrix that has a Hankel structure. For future reference, we introduce the following SVD 
\begin{equation}
\H_{\y} = \matQ\Sigmab\matP.
\label{Eq:SVDHy}
\end{equation} 
In the sequel, we consider model order selection schemes that use $\H_\y$ to estimate an appropriate order.  
	
\section{Model Order Selection Rules}\label{sec:review}

Model order selection techniques may estimate the dimension of the signal space using statistical information about the noisy observations \cite{Stoica2004}. Also, when the signal model satisfies \eqref{Eq:Signal_1}, we can benefit from its particular algebraic structure, as in \eqref{Um:eq} \cite{Badeau2006,Papy2007}. In this section, we review three different techniques for model order selection. The first two are only suitable for models that satisfy the rotational invariance property. On the other hand, the third one only considers the nuisance of the random perturbation onto the signal. 
	
	\subsection{Algebraic Structure of the Signal}

Suppose that $\H_\x$ is available, i.e., we are in the noiseless case.  Define the matrix $\matU(s) \in\C^{m\times s}$ that 
        contains the first $s$ columns of $\matU$, and $\matU_f(s)$ and $\matU_l(s)$ following  \eqref{Eq:matrices_lf}. When $s=r$,  $\matU_f(s)$ and $\matU_l(s)$ span the same subspace according to \eqref{Um:eq}.  The  closeness between the column spaces of $\matU_f(s)$ and $\matU_l(s)$ provides a key to estimate $r$. We use the principal angles as a measure of  proximity between subspaces.

		\begin{definition}\label{def1}
			Let $\setX,\setY \subseteq\C^n$ be complex subspaces with $\dim(\setX)=\dim(\setY) = s$. The principal angles 	between $\setX$ and $\setY$ are 
			\begin{equation*}
				\Thetab(\setX,\setY) = \big[\theta_1,\ldots,\theta_s\big], \quad \theta_k\in[0,\pi/2],\quad k= 1,\ldots,s
	            %\label{Eq:principalAngles}		 
			\end{equation*}
			which are  recursively defined by
	    	\begin{equation}
	    		\begin{aligned} 
					\cos\theta_k & =\frac{|\langle\x_k,\y_k\rangle|}{\|\x_k\|_2\|\y_k\|_2}\\ &= \max_{\stackrel{\x\in\setX}{\y\in\setY}}\bigg\{ \frac{|\langle\x,\y\rangle|}{\|\x\|_2\|\y\|_2} &:\ \x^H\x_i=0,\y^H\y_i=0,\\ & &\forall i\in\{1,\ldots,k-1\}\bigg\}.
				\end{aligned}
				\label{Eq:principalAngles2}
			\end{equation}
			The vectors $\{\x_1,\ldots,\x_s\}$ and $\{\y_1,\ldots,\y_s\}$ are called the principal vectors.
	  	\end{definition}
	
        \begin{lemma}\label{Lemma:0}
            $\setX=\setY$, if and only  $\Thetab(\setX,\setY) = 0$
        \end{lemma}
        \begin{proof}
            By construction, the principal vectors $\{\x_1,\ldots,\x_s\}$ are linearly independent vectors that span the subspace $\setX$ because $\dim(\setX)= s$. Similarly, the principal vectors $\{\y_1,\ldots,\y_s\}$ span $\setY$. If $\cos\theta_k = 1$, we have that $\x_k$ and $\y_k$ are aligned. Therefore, if all the principal angles are zero, the sets $\{\x_1,\ldots,\x_s\}$ and $\{\y_1,\ldots,\y_s\}$ generate the same subspace. 

            On the other hand, when $\setX = \setY$, it is clear that   
            \[
                \max_{\stackrel{\x\in\setX}{\y\in\setY}} \frac{|\langle\x,\y\rangle|}{\|\x\|_2\|\y\|_2} = 1
            \]
            and therefore, $\cos \theta_k = 1$ for all $k$. 
        \end{proof}

\begin{definition}
 The gap distance between  $\setX$ and $\setY$ is given by 
        \begin{equation}
	        \rho(\setX,\setY) = \max\bigg\{\max_{\stackrel{\x\in\setX}{\|\x\|_2=1}}\mathrm{dist}(\x,\setY),\max_{\stackrel{\y\in\setY}{\|\y\|_2=1}}\mathrm{dist}(\y,\setX)\bigg\},
	        \label{Eq:gapSubspaces}
        \end{equation}
\end{definition}

 \begin{prop}\label{Prop:gap}
    Let $\setX$ and $\setY$ be two subspaces, and denote $\matProj_\setX$ and $\matProj_\setY$ the orthogonal projections onto $\setX$ and $\setY$ respectively. Then
    \begin{itemize}
	    \item[i)] $\rho(\setX,\setY) =     \|\matProj_\setX-\matProj_\setY\|_2$;
	    \item[ii)] $\rho(\setX,\setY) = \sin\theta_1$, with $\theta_1$ the maximum principal angle between $\setX$ and $\setY$.
    \end{itemize}
\end{prop}
\begin{proof}
The proof of this proposition is in  \cite[Th.4.5]{Stewart90}. 
\end{proof}

Let $\matProj_f(s)$ and $\matProj_l(s)$ be the projection matrices onto the column spaces of $\matU_f(s)$ and $\matU_l(s)$ respectively. Define $\theta_1(s)\ge\cdots\ge\theta_s(s)$ as the principal angles between the $s$-dimension column-spaces of $\matU_f(s)$ and $\matU_l(s)$. According to Proposition \ref{Prop:gap}, the gap distance between $\matU_f(s)$ and $\matU_l(s)$ is 
\begin{equation}
    \rho(s) = \|\matProj_f(s) - \matProj_l(s)\|_2 = \sin \theta_1(s)
    \label{eq:gapFunction}
\end{equation}

%From lemma \ref{Lemma:0} we have that $0\le\rho(s)\le 1$ for all $1\le s\le m-1$, and   $\rho(r)=0$  only when $s=r$.

\begin{theorem}\label{The1:ESTER_GAP}
    Consider the function 
    \begin{equation}
        \matE(s) = \matU_f(s) - \matU_l(s)\matU_l(s)^\dagger\matU_f(s),
        \label{eq:ESTER_function}
    \end{equation}
    Then $\|\matE(s)\|_2 \le \sin\theta_1(s).$
\end{theorem}
\begin{proof}
Consider the polar decomposition of $\matU_f(s)$ and $\matU_l(s)$
\begin{equation}
    \begin{aligned}
        & \matU_f(s) = \hat{\matU}_f(s)\big(\matI_s -         \u_f^H\u_f\big)^{\frac{1}{2}}\\
	    & \matU_l(s) = \hat{\matU}_l(s)\big(\matI_s -         \u_l^H\u_l\big)^{\frac{1}{2}}
	\end{aligned}
    \label{Eq:nearest_Orthonormal}
\end{equation}
where  $\u_f\in\C^{1\times s}$ ($\u_l\in\C^{1\times s}$) is the first (last) row of $\matU(s)$, and $\hat{\matU}_f(s)$ and $\hat{\matU}_l(s) $ are $(m-1)\times s$-complex matrices, both with  orthonormal columns. The orthogonal projections onto the column spaces of $\matU_f(s)$ and $\matU_l(s)$ are:  
\begin{equation}
    \begin{aligned}
    & \matProj_f(s) = \matU_f(s)\matU_f(s)^\dagger = \hat{\matU}_{f}(s)\hat{\matU}_{f}^H(s) \\[0.3em]
    & \matProj_l(s) = \matU_l(s)\matU_l(s)^\dagger = \hat{\matU}_{l}(s)\hat{\matU}_{l}^H(s),
    \end{aligned}
    \label{Eq:ProjMatrix}
\end{equation}
where $\matU_l(s)^\dagger$ is the Moore-Penrose pseudo-inverse. Now, let  
 \begin{equation}
    \hat{\matE}(s) = \hat{\matU}_f(s) - \hat{\matU}_l(s)\hat{\matU}_l(s)^H\hat{\matU}_f(s).
    \label{eq:ESTER_gap}
\end{equation}
Since $\hat{\matU}_f$ and $\hat{\matU}_l$ have orthonormal columns
\[\begin{aligned} \|\hat{\matE}(s)\|_2 & = \|\hat{\matU}_{f}(s) -\hat{\matU}_{l}(s)\hat{\matU}_l^H(s)\hat{\matU}_f(s)\|_2 \\[0.3em] & = \|\left(\hat{\matU}_{f}(s)\hat{\matU}_{f}^H(s) -\hat{\matU}_{l}(s)\hat{\matU}_l^H(s)\right)\hat{\matU}_f(s)\|_2\\[0.3em]
&= \|\matProj_f(s) - \matProj_l(s)\|_2 = \rho(s) \end{aligned}
 \]
 where we have used Proposition \ref{Prop:gap} in the last equality. Now, 
 \begin{equation}
  	\begin{aligned} \|\matE(s)\|_2 & = \|\matU_{f}(s) -\matU_{l}(s)\matU_l(s)^\dagger\matU_f(s)\|_2 \\[0.3em] & = \|\left(\matU_{f}(s)\matU_{f}(s)^\dagger -\matU_{l}(s)\matU_l(s)^\dagger\right)\matU_f(s)\|_2\\
    	&\le \|\matProj_f(s) - \matProj_l(s)\|_2\|\matU_f(s)\|_2  \\[0.3em]
    	&= \rho(s)\|\matU_f(s)\|_2. \end{aligned}
    	\label{Eq:ESTER_gap_Theo}
 \end{equation}
Since $\matU_f(s)$ is a submatrix of the unitary matrix $\matU$, its singular values will be at most 1. Then $\|\matU_f(s)\|_2\leq 1$ and the result follows. 
\end{proof}

%On the other hand, we can write $\matE(r) = (\matI_{m-1}-\matProj_l)\matU_f(r)$, where $\matI_{m-1}-\matProj_l$ is the projector onto the orthongonal complement of the columns space of $\matU_l(r)$ and we have that $(\matI_{m-1}-\matProj_l)\matU_f(r)=\mathbf{0}$. As a result we have that $\rho(r) = \|\matE(r)\|_2=0$. 

\begin{theorem}\label{The1:SAMOS}
        Define the augmented matrix
      \begin{equation}
	    \matE_{aug}(s)=[\matU_f(s) \,\,\,\, \matU_l(s)].
	    \label{Eq:SAMOS_augMatrix}
        \end{equation}        
 Let  $\gamma_1 \ge \cdots \ge \gamma_{2s}$ be its  singular values. Then,
\begin{equation}
    \frac{1}{s}\sum_{i=s+1}^{2s}\gamma_i \le \sqrt{2} \left[1+ \frac{1}{s}\sum_{i=1}^{s}\sin\frac{\theta_{i}(s)}{2}\right]
    \label{eq:Angle_Weyl}
    \end{equation}
%    for $i = s+1,\ldots, 2s.$
\end{theorem}
\begin{proof}
Recalling that  $\matU_f(s)$ is a full rank matrix, we have that $\matU_f(s)$ and $\hat{\matU}_f(s)$ span the same column space, and likewise $\matU_l(s)$ and $\hat{\matU}_l(s)$. Then, the principal angles between $\matU_f(s)$ and $\matU_l(s)$ are the same as the angles between  $\hat{\matU}_f(s)$ and $\hat{\matU}_l(s)$. 
Define
\begin{equation} 
    \hat{\matE}_{aug}(s)=[\hat{\matU}_f(s) \,\,\,\, \hat{\matU}_l(s)].
\end{equation}
Notice that the singular values of $\hat{\matE}_{aug}(s)$ are obtained from the matrix $\matI_{2s} + \mathbf{M}$, where
\[\mathbf{M} = \begin{bmatrix} \mathbf{0} & \hat{\matU}_f(s)^H\hat{\matU}_l(s)\\[.3em] \hat{\matU}_l(s)^H\hat{\matU}_f(s) & \mathbf{0} 
\end{bmatrix}.\]

According to  \cite[Th. I.5.2]{Stewart90}, 
the singular values of $\hat{\matU}_f(s)^H\hat{\matU}_l(s)$ arranged in non-increasing order are $\cos\theta_1(s),\ldots,\cos\theta_s(s)$. Then, from \cite[Th. 7.3.3]{Horn1990}, the last $s$ singular values of $\hat{\matE}_{aug}$ arranged in non-increasing order are
\[\sqrt{1-\cos\theta_{2s-i+1}(s)} = \sqrt{2}\sin\frac{\theta_{2s-i+1}(s)}{2} \quad i = s+1, \cdots ,2s.\]
Let  $\matA = \matU_{f}(s) - \hat{\matU}_f(s)$, and $\matB = \matU_{l}(s) - \hat{\matU}_l(s)$. Therefore, $\matE_{aug}(s) - \hat{\matE}_{aug}(s) =  [\matA\ \matB]$.
It was proven in \cite{Higham89} that $\hat{\matU}_f(s)$ is the nearest matrix  to $\matU_f(s)$  with  orthonormal columns, and so is $\hat{\matU}_l (s)$ with $\matU_l(s)$. Moreover,
\[\|\matA\|_2 = \max_i|\zeta_i-1|,\]
where $\zeta_i$ is the $i$-th largest singular value of $\matU_{f}(s)$. Notice that $\zeta_i\leq 1$ because $\matU_{f}(s)$ is a submatrix of the unitary matrix $\matU$.  Therefore, $\|\matA\|_2  \le 1$, and similarly, $\|\matB\|_2\le 1$. According to Weyl Theorem \cite[Th. 4.11]{Stewart90}, the last $s$ singular values of $\matE_{aug}(s)$ and $\hat{\matE}_{aug}(s)$ are separated at most by $\|\matA \,\,\,\, \matB\|_2$, i.e., for $i=s+1, \cdots, 2s$,
	\begin{equation}
	    \begin{aligned} 
	     |\gamma_i-\sqrt{2}\sin\frac{\theta_{2s+1-i}(s)}{2}|& \le \|[\matA \,\,\,\, \matB]\|_2  \\
	     & \le \sqrt{\|\matA\|_2^2+\|\matB\|_2^2}\leq \sqrt{2}.
	     \end{aligned}
    \label{Eq:GammaTh2}
	\end{equation}
   Using the triangle inequality we get
	\[ \begin{aligned} 
	 & \frac{1}{s}\sum_{i=s+1}^{2s}\gamma_i-\frac{1}{s}\sum_{i=s+1}^{2s}\sqrt{2}\sin\frac{\theta_{2s+1-i}(s)}{2} \leq \\[0.3em]
	 & \le \frac{1}{s}\sum_{i=s+1}^{2s}
	\bigg|\gamma_i-\sqrt{2}\sin\frac{\theta_{2s+1-i}(s)}{2}\bigg|\le \sqrt{2}.\end{aligned}\]
Then,
	\[  \frac{1}{s}\sum_{i=s+1}^{2s}\gamma_i\leq \sqrt{2}+ \frac{1}{s}\sum_{i=1}^{s}\sqrt{2}\sin\frac{\theta_{i}(s)}{2},
	\]
where we have rearranged the terms in the last summation. 
\end{proof}

\begin{figure}[b]
    \begin{subfigure}{.45\linewidth}
	    \centering
	    \includegraphics[height=0.75\textwidth]{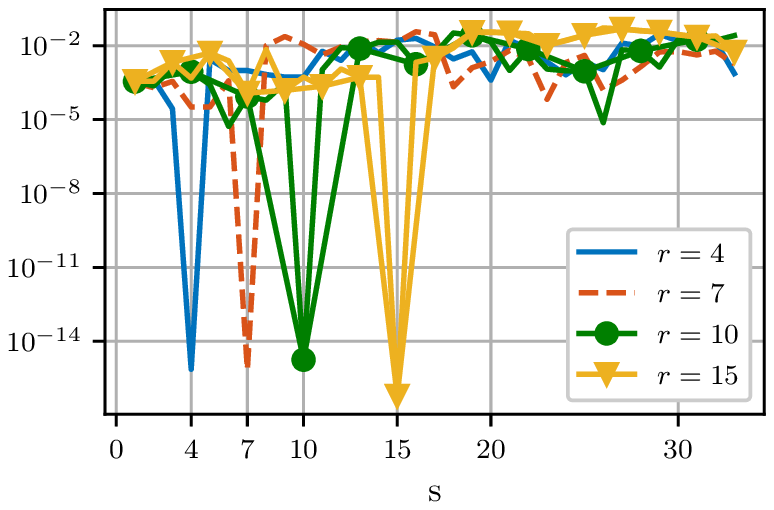}
		%\centerline{\resizebox{\textwidth}{0.9\linewidth}{\input{img/bounds_ESTER_function.pgf}}}
		%\vspace{-0.3cm}
		\caption{$|\|\matE(s)\|-\rho(s)|$ %Theorem \ref{The1:ESTER_GAP}
		}%\medskip
	\end{subfigure}
	~
	\begin{subfigure}{.45\linewidth}
	    \centering
	    \includegraphics[height= 0.75 \textwidth]{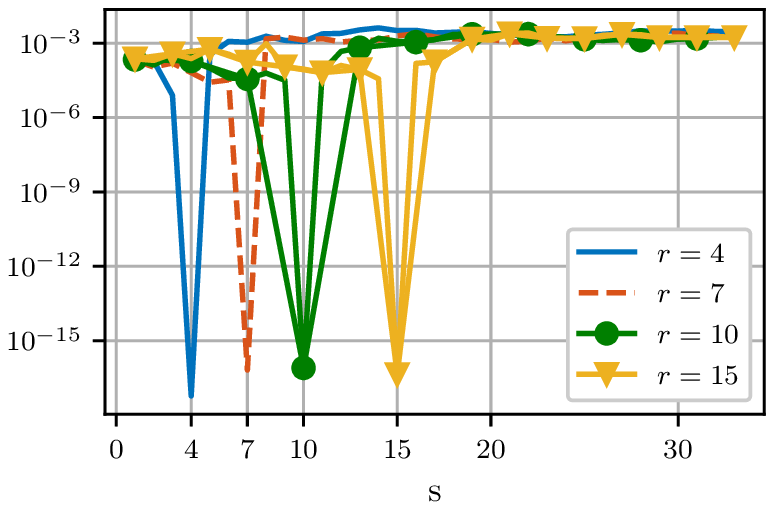}
		%\centerline{\resizebox{\textwidth}{0.9\linewidth}{\input{img/bounds_SAMOS_function.pgf}}}
		%\vspace{-0.3cm}
		\caption{$|\sum_i(\gamma_{s+i}-\sqrt{2}\sin(\theta_{i}/2))|$
		%Difference between the two functions in Theorem \ref{The1:SAMOS}
		}%\medskip
	\end{subfigure}
	\caption{Evaluation of the bounds given in Th. \ref{The1:ESTER_GAP} and \ref{The1:SAMOS}.}
	\label{Fig:Bounds_ESTER_SAMOS}
\end{figure}

We have shown experimentally that both Theorems \ref{The1:ESTER_GAP} and \ref{The1:SAMOS} provide tight upper bounds. For that, we have simulated  \eqref{Eq:Signal_1} using different values for $r$. For each signal, the frequencies $z_i=e^{2\pi\jmath f_i}$ were selected by taking $r$ frequencies uniformly spread in the interval $(0,1]$. The complex amplitudes $c_i$ were independent samples of the uniform distribution in $[1,1.5]$.
%In order to show the that the upper bounds found in Theorems \ref{The1:ESTER_GAP} and \ref{The1:SAMOS} are tight, we compute the functions $\|\matE(s)\|_2$, the singular values of $\matE_{aug}(s)$ and the principal angles, from signals defined in \eqref{Eq:Signal_1} with  different orders $r$. We consider $z_i=e^{2\pi\jmath f_i}$ by taking $r$ frequencies in $(0,1]$ with equal mutual separation $1/r$ and the amplitudes $c_i$ are selected in such way that their real and imaginary parts are drawn from an i.i.d uniform random variable in $[1,1.5]$. Fig. \ref{Fig:Bounds_ESTER_SAMOS} shows the absolute value of the difference between the functions involved in theorems \ref{The1:ESTER_GAP} and \ref{The1:SAMOS}. 
Fig. \ref{Fig:Bounds_ESTER_SAMOS} shows the 
results as $s$ varies. To assess the bound in Th. \ref{The1:ESTER_GAP},  Fig.\ref{Fig:Bounds_ESTER_SAMOS}(a) shows   $|\|\matE(s)\|_2-\rho(s)|$. Th. \ref{The1:SAMOS} is analyzed in Fig. \ref{Fig:Bounds_ESTER_SAMOS}(b) that displays $1/s|\sum_{i=1}^s(\gamma_{s+i}-\sqrt{2}\sin(\theta_{i}(s)/2))|$. Both figures show that the bounds are tight for all $s$. Moreover, they are both minimized when $s=r$.

When $s=r$, the column spaces of $\matU_f(s)$ and $\matU_l(s)$ are aligned. Then,  all the angles $\theta_i(s)$, $i=1, \ldots, s$,  are equal to zero. Therefore, $\rho(s)=0$ and we obtain the right order by minimizing $\|\matE(s)\|_2$. Also in this case, $\rank(\matE_{aug}(s))=r$, $\gamma_{s+1} = \cdots = \gamma_{2s}=0$, and minimizing \eqref{eq:Angle_Weyl} is also a good alternative. 
However, both observations rely on the knowledge of matrix $\matU$, which is only possible in the noiseless case.

%We have shown experimentally that $\rho(s)$ is a tight upper bound to $\|\matE(s)\|_2$. When $s=r$, $\rho(r) = 0$. Then, by minimizing $\|\matE(s)\|_2$ we obtain the right order model. Clearly, this is true only in the noiseless case,  when $\matU$ is available.

 Since $\H_x$ is only observed thru $\H_y$, $\matU$ is not directly known, and we have to work with $\matQ$ and its submatrices instead. Using these matrices, the order estimation rule known as ESTimation Error Rule (ESTER) was proposed in  \cite{Badeau2004}. The rule minimizes the function
\begin{equation}
    J_{ESTER}(s) = \|\matQ_f(s) - \matQ_l(s)\matQ_l(s)^\dagger\matQ_f(s)\|_2.
    \label{eq:rhatESTER}
\end{equation}
 An alternative approach,  the subspace-based automatic model order selection (SAMOS), was proposed in \cite{Papy2007}. In this case, the rule selects the order that minimizes the sum of $\varsigma_i$, $i=s+1, \ldots, 2s$, which are the last singular values of $[\matQ_f(s) \,\,\,\, \matQ_l(s)]$
 \begin{equation}
    J_{SAMOS}(s)=\frac{1}{s}\sum_{i=s+1}^{2s} \varsigma_i.
	        \label{eq:rhatSAMOS}
\end{equation}

%When noise is present the rotational invariance principle in \eqref{Um:eq} is no longer valid, due to we have to work with matrix $\matQ$ from the SVD of $\H_{\y}$. However, a suitable truncation of $\matQ$ may approximate equality in \eqref{Um:eq}. In this case, we work with the subspaces generated by the the columns of $\matQ_l(s)$ and $\matQ_f(s)$. The problem here is that when $s=r$ the two subspaces are not going to be completely aligned, this means that the angles are not zero, but will be subject to perturbation. 

Let $\vartheta_i(s), i=1, \cdots, s$ be the principal angles between $\matQ_l(s)$ and $\matQ_f(s)$ ordered in non-increasing order. Th. \ref{The1:ESTER_GAP} and \ref{The1:SAMOS} state upper bounds for the cost used in  the ESTER and SAMOS methods. Notably,
\begin{equation}
    J_{ESTER}(s) \leq \sin\vartheta_1(s).
\end{equation}
%On the hand, Theorem \ref{The1:SAMOS} gives an upper bound for the cost function used in SAMOS
     \begin{equation}
        J_{SAMOS}(s) \le \sqrt{2} \left[1+ \frac{1}{s}\sum_{i=1}^{s}\sin\frac{\vartheta_{i}(s)}{2}\right].
    \label{eq:SamosBoundHy}    
     \end{equation}

It was shown in \cite{Badeau2006, Papy2007} that these methods outperform information-theoretic criteria such as AIC or MDL. However, these techniques do not have good performance under high noise level. Notice that when $s$ is close to $r$ the angles $\vartheta_i(s), i=1, \ldots ,s$ are small, and $\sin(\vartheta_i(s))$ is very sensitive to small deviations. As a consequence, both techniques have a poor performance when the noise level is high as it was observed experimentally. Although in SAMOS the average shown in \eqref{eq:SamosBoundHy} may reduce the effect of noise, this is not completely effective when we are dealing with low signal to noise ratios.  In section \ref{sec:results} we show some numerical experiments that support these claims. 

	\subsection{Statistical Structure of the Noise}
A different approach for model order estimation is to infer  	$\mathrm{rank}(\H_\x)$ by counting the relevant singular values of $\H_\y$.  When given a hard threshold $\tau$, the model order is estimated as
		\begin{equation}
			\hat{r}_{thr} = \left|\{\sigma_i : \sigma_i>\tau\}\right|.
			\label{eq:rank1}
		\end{equation}
	Here $|.|$ is the size of the set. The choice of $\tau$ is a delicate matter. A large value for $\tau$ could result in selecting a rank lower  than desired. On the other hand, a small $\tau$ leads to overestimating the model order. This problem was studied in \cite{Gavish2014}, where the authors considered a perturbation matrix with independent identically distributed (i.i.d.) entries. Under an  appropriate asymptotic framework, the authors obtain that the optimal value for $\tau$ is 
	%\textcolor{red}{Poner la expresi'on para matrices rectangulares y para cuadradas}
		\begin{equation}
			\tau =  \kappa\sqrt{m}\eta
			\label{eq:Gavish_Th}   
		\end{equation}
		where $\eta$ is the white noise level and $\kappa$ is a constant that depends on the matrix dimensions 
		\begin{equation*}
			\kappa = \sqrt{2(c+1)\frac{8c}{c+1+\sqrt{c^2+14c+1}}}.
		\end{equation*}

\begin{figure}[t]
		    \begin{subfigure}[b]{0.45\linewidth}
		        \centering
		        \includegraphics[width =  \textwidth]{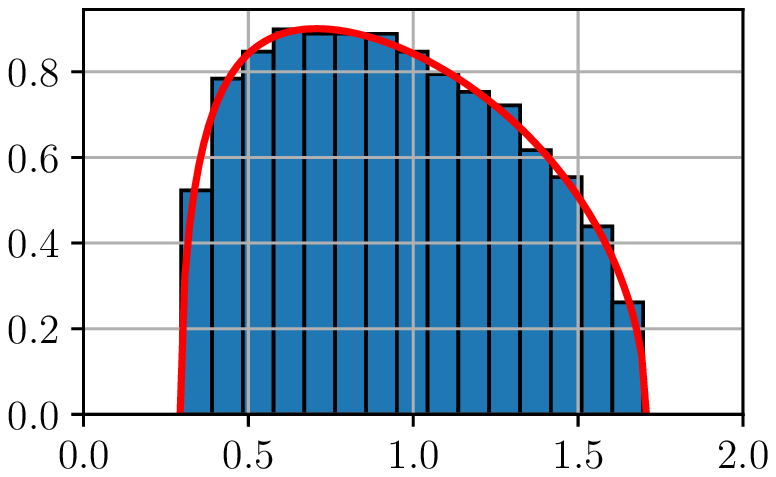}
		        \caption{Random matrix with i.i.d. entries.}
		        \label{Fig:svd1_iid}
		    \end{subfigure}
		    ~
		    \begin{subfigure}[b]{0.45\linewidth}
		        \centering
		        %\resizebox{\textwidth}{0.8\linewidth}{\input{img/singular_values_Hankel_entries.pgf}}
		        \includegraphics[width = \textwidth]{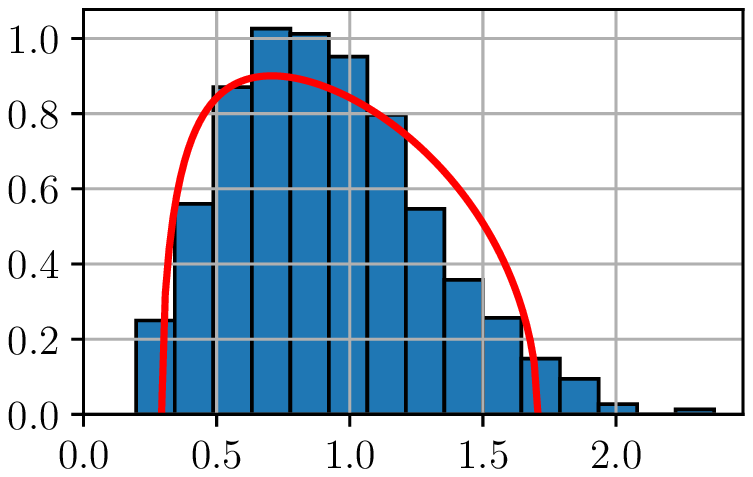}
		        \caption{Random matrix with Hankel structure.}
		        \label{Fig:svd1_Hankel}
		    \end{subfigure}
		    \caption{Normalized histograms of singular values (blue) and the quarter circle density (red).}
		 	\label{Fig:svd1}
		  \end{figure}

The result follows from the limiting distribution of the singular values of a matrix with i.i.d. entries. Let $\matW$ be a $m\times n$ matrix  ($m\ge n$)  with i.i.d. entries. Denote $c = \frac{n}{m}$. Then the empirical distribution of the singular values of   $\matW/\sqrt{m}$ follows  the Marchenko-Pastur density%\textcolor{red}{the quarter circle density... ésta no es la MP density?}
		\begin{equation}
			f_{MP}(x) = \frac{\sqrt{4c-(x^2-1-c)^2}}{\pi c x}\cdot\mathbf{1}\big\{x\in[c_{-},c_{+}]\big\}.
			\label{Eq:MP_density}
		\end{equation}
		with $c_{\pm} = 1\pm\sqrt{c}$. In Fig.~\ref{Fig:svd1_iid} we show the histogram of the singular values of $\mathbf{W}/\sqrt{m}$ using Gaussian entries with zero mean and unit variance, when $m=1024, n =512$. We have also plot $f_{MP}(x)$. It follows that the largest singular value due to noise is approximately $1+\sqrt{c}$. Thus, singular values associated to the signal that are below this threshold will not be differentiated from those of noise.
		
		When the perturbation matrix has i.i.d. entries, experimental results have shown that  threshold \eqref{eq:Gavish_Th} leads to  good performance, even when the SNR is negative. However, when considering a sum of exponentials, the perturbation matrix $\H_{\w}$ inherits the Hankel structure and its entries are not i.i.d.  In this case, the empirical spectral distribution converges to a nonrandom symmetric probability measure which has no explicit expression \cite{Bryc2006}. Fig.~\ref{Fig:svd1_Hankel} shows the histogram of the  singular values of a matrix $\H_{\w}/\sqrt{m}$ with $m=1024, n=512$.
		In this case, there are singular values that fall outside the support of $f_{MP}(x)$, and choosing a threshold following \eqref{eq:Gavish_Th} may lead to poor performance. 
	
\section{Random Hankel matrix}\label{sec:RandomHankel}

Let $\lambda_i$ and $\sigma_i$, $i=1, \cdots , n$, be the singular values of $\H_x$ and $\H_y$ respectively arranged in non-decreasing order.  Following Weyl's Theorem  we have that 
\begin{equation}
		|\sigma_i - \lambda_i|\le \|\H_{\y} - \H_{\x}\|_2 = \|\H_{\w}\|_2
		\label{eq:Weyl}
	\end{equation}
Since $\rank(\H_{\x}) = r$,  $\lambda_{i} = 0$ if $i > r$,. Then
		\begin{equation}
			\sigma_n \le \sigma_{n-1}\le \cdots \le \sigma_{r+1} \le \|\H_{\w}\|_2.
			\label{eq:Weyl2}
		\end{equation} 
%		Then, all the singular values associated with the noise subspace are less or equal to the spectral norm of the Hankel matrix associated with the noise vector. Because by construction the spectral norm of the Hankel matrix $\H_{\w}$ is a random variable, we will upper bound it in probability. In order to find this bound we state the following two results.

\begin{lemma}\label{Lemma:1}
	Consider the complex vector $\w \in\C^{m+n-1}$. Let $\H_{\w}$, be the $m\times n$ Hankel matrix associated with $\w$. %, and  $\matC_{\w}$, be the  $(m+n-1)\times (m+n-1)$ circulant matrix associated with $\w$. 
	Then		
	\begin{equation}
		\|\H_{\w}\|_2 \leq %\|\matC_{\w}\|_2 =
		\max_{0\leq k\leq m+n-2}|\mathbf{e}_{k+1}^T\matV\w|,
		\label{Eq:CircMatrixNorm}
	\end{equation}
	where $\matV$ is the DFT matrix and $\mathbf{e}_k$ is the $k$-th unitary vector. 
\end{lemma}
\begin{proof}
	See appendix \ref{Proof:Lemma1}
\end{proof}
													
%When the circulant matrix has complex Gaussian random entries, we characterize the spectral norm of the circulant random matrix with the following distribution function 

\begin{lemma} \label{Lemma:2}
If $\w\sim\mathcal{CN}(\mathbf{0},\eta^2\matI_{m+n-1})$, then 
		\begin{equation}
		    \P[ \|\H_{\w}\|_2 \leq \tau] \geq \bigg[1-e^{-\frac{\tau^2}{(m+n-1)\eta^2}}\bigg]^{m+n-1} \cdot\one\big\{\tau\geq 0\big\}.
		    \label{Eq:BoundCirculantMatrix2}
		\end{equation}		 
\end{lemma}
\begin{proof}
    See Appendix \ref{Proof:Lemma2}
\end{proof}
		
\begin{theorem}\label{Th:boundHankelMatrix2}
    Let $\H_{\w}$ be a $m\times n$ random Hankel matrix with generating vector $\w \sim\mathcal{CN}(\mathbf{0},\eta^2\matI_{m+n-1})$. Then, for any $\beta\in [0,1]$, we have that
    \[
    \P[\|\H_{\w}\|_2\le \tau_1]\ge\beta.
    \]
    where 
	\begin{equation}
	   \begin{aligned} 
	      \tau_1 & =  \sqrt{-(m+n-1)\eta^2\log(1-\beta^{\frac{1}{m+n-1}})}. 
	    \end{aligned}
			\label{Eq:BoundHankelMatrix2}
	\end{equation}
\end{theorem}				
\begin{proof}
The result follows from Lemma~\ref{Lemma:2} by selecting $\beta$ so that
\[
\beta \leq \bigg[1-e^{-\frac{\tau_1^2}{(m+n-1)\eta^2}}\bigg]^{m+n-1} 
\]
%	From lemma~\ref{Lemma:2} we obtain that with probability greater than $\beta$ 
%	\[\|\H_{\w}\|_2\le \sqrt{-(m+n-1)\eta^2\log(1-\beta^{\frac{1}{m+n-1}})}.\]
%	Then, by lemma~\ref{Lemma:1} we obtain \eqref{Eq:BoundHankelMatrix2}.
\end{proof}
		
When dealing with real random Hankel matrices, Lemma~\ref{Lemma:2} is no longer valid, and we cannot follow the same path to obtain a bound on $\|\H_\w\|_2$. A possible solution is to perform Montecarlo simulations of the spectral norm of the Hankel matrix to obtain an empirical bound as in \cite{Albert2021}. To avoid lengthy simulations, here we propose a workaround by using concentration inequalities. 
\begin{theorem}
\label{Th:HankelReal}
 	Let $\H_{\w}$ be a $m\times n$ random Hankel matrix with generating vector $\w \sim\mathcal{N}(\mathbf{0},\eta^2\matI_{m+n-1})$. Then, for any $\beta \in [0,1]$ we have that 
 	\[
 	 \P[\|\H_{\w}\|_2\le \tau_2]\ge\beta.
 	\]
 	where
 		 \begin{equation}
 		    \begin{aligned} 
		        %\|\H_{\w}\|_2 \leq & 
		        \tau_2 = \sqrt{-2\max\{m,n\}\eta^2\log\frac{1-\beta}{(m+n)}} %\\[0.3em] & \text{with prob. $\beta$}
		    \end{aligned}
		    \label{Eq:BoundHankelMatrix}
		\end{equation}
\end{theorem}		
\begin{proof}
Consider the following concentration inequality  \cite[chap. 4]{tropp2015}
		\begin{equation} 
		    \P[\|\H_{\w}\|_2\geq \tau] \leq (m+n) \exp\bigg[-\frac{\tau^2}{2\eta^2\max\{m,n\}}\bigg].
		\label{Eq:InequalityTroop}
		\end{equation}
Then, the result follows by taking 

$$\beta \leq 1- (m+n) \exp\bigg[-\frac{\tau^2}{2\eta^2\max\{m,n\}}\bigg]$$.
\end{proof}	

A bound similar to \eqref{Eq:BoundHankelMatrix} was obtained in \cite{Qiao2020} for the case of real square Toeplitz matrices with gaussian elements. The bound we have just derived also works for rectangular matrices. 
%
%In case of real random Hankel matrix, the bound found in \cite{Qiao2020} and the concentration inequalities in \cite{tropp2015} we obtain

\begin{figure}[t]
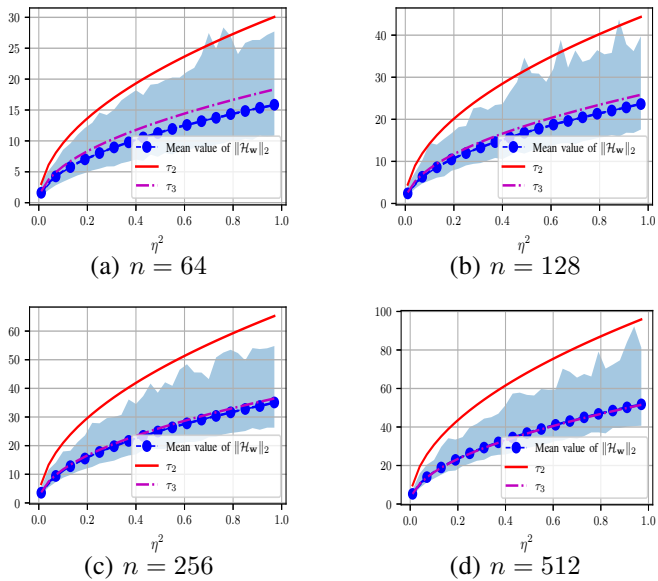

	\begin{minipage}[b]{.45\linewidth}
	    \centering
		 \centerline{\resizebox{\textwidth}{0.9\linewidth}{\input{img/taus_N_64_beta_0.9.pgf}}}
		 \vspace{-0.3cm}
		 \centerline{(a) $n=64$}\medskip
	\end{minipage}
	\hfill
	\begin{minipage}[b]{.45\linewidth}
		\centering
		\centerline{\resizebox{\textwidth}{0.9\linewidth}{\input{img/taus_N_128_beta_0.9.pgf}}}
		\vspace{-0.3cm}
		\centerline{(b) $n=128$}\medskip
	\end{minipage}
	
	\begin{minipage}[b]{0.45\linewidth}
		\centering
		\centerline{\resizebox{\textwidth}{0.9\linewidth}{\input{img/taus_N_256_beta_0.9.pgf}}}
		\vspace{-0.3cm}
		\centerline{(c) $n=256$}\medskip
	\end{minipage}
	\hfill
	\begin{minipage}[b]{0.45\linewidth}
		\centering
		\centerline{\resizebox{\textwidth}{0.9\linewidth}{\input{img/taus_N_512_beta_0.9.pgf}}}
		\vspace{-0.3cm}
		\centerline{(d) $n=512$}\medskip
	\end{minipage}
	\caption{Different realization of real and square $\|\H_\w\|_2$ and bounds  $\tau_2$, and $\tau_3$ for different  dimensions $n$ with $\beta = 0.9$}
\label{fig:taus11}
		 \end{figure}

Bounds $\tau_1$ and $\tau_2$ establish hard thresholds that may be used as  \eqref{eq:Gavish_Th} to separate the signal space from the noise space. To compare these bounds, we have considered the case of real signals and square Hankel matrices, i.e., $m=n$. Since we are dealing with real matrices, we consider $\tau_2$ as in \eqref{Eq:BoundHankelMatrix}. %In order to find a bound as in Theorem \ref{Th:boundHankelMatrix2} for real random vector, we find the quantile such as
%\[\P[\|\matC_{\w}\|_2\le\tau_1] = \beta, \]
%with distribution function in \eqref{Eq:DistributionRealVector}. 
For square matrices, we compute the bound in \eqref{eq:Gavish_Th} as
\begin{equation}
	\tau_3 = \frac{4}{\sqrt{3}}\sqrt{n}\eta.
	\label{eq:gavish}
\end{equation}
For each realization of the random vector $\w\sim\mathcal{N}(\mathbf{0},\eta^2\matI_{2n-1})$, we compute the spectral norm of the associated square Hankel matrix. 
Fig.~\ref{fig:taus11} shows the behavior of $\|\H_\w\|_2$ as the noise level $\eta^2$ increases. Each subplot corresponds to a different value of $n$. The shaded area  represents the dispersion  among the realizations of $\|\H_\w\|_2$ together with  the bounds  $\tau_2$ and $\tau_3$. 

As the matrix dimension increases, $\tau_3$ approaches the mean value of $\|\H_\w\|_2$. As a consequence, when we estimate the order with $\tau_3$ we take into account singular values associated with the noise subspace, as it was also observed in Fig.~\ref{Fig:svd1_Hankel}. On the other hand, from Fig.~\ref{fig:taus11} we see that $\tau_2$ is a conservative bound, so some singular values corresponding to the signal subspace may fall under this threshold. %To overcome this problem, it is possible to perform Montecarlo simulations of the spectral norm of the Hankel matrix in order to obtain an empirical bound. In \cite{Albert2021} an empirical expression that depends on the matrix dimension was found by performing Montecarlo simulations. 
To overcome this problem, in the next section we formulate a constrained optimization problem.

\section{Constrained Selection Rule}\label{sec:SelectionRule}

    We have observed that model order selection rules based on the singular value distribution of $\H_\y$ lose performance when strong signals are present because they do not consider the algebraic structure of the Hankel matrix. On the other hand, rules like ESTER or SAMOS underperform when the SNR is low because they are built on noise-free assumptions.

	Here, we propose to combine both approaches in a single optimization problem to overcome both problems. In particular we impose a restriction on the singular values associated with the noise subspace. From inequality \eqref{eq:Weyl2} we have
	\begin{equation}
		 \sigma_{s+1} \le \|\H_{\w}\|_2 \le \tau
		  \label{Eq:BoundHankelMatrix3}
	\end{equation}
	with probability $\beta$. Notice that Theorems \ref{Th:boundHankelMatrix2} and \ref{Th:HankelReal} give appropriate upper bounds for the singular values associated with the noise subspace. Based on these observations, we propose the following constrained optimization problem 
	\begin{equation}
		\begin{aligned}
			\hat{r}_{c} = \arg  & \min_{s\in \N} \mathcal{J}(s)\\
			& \text{s.t. }  \sigma_{s+1}<\tau,\text{ with probability } \beta.
		\end{aligned}
		\label{eq:propuesta3}
	\end{equation}
	where $\mathcal{J}(s)$ can be either $J_{ESTER}(s)$ or $J_{SAMOS}(s)$ and $\tau$  is defined in Theorem \ref{Th:boundHankelMatrix2} or \ref{Th:HankelReal} whether the noise is complex or real. The heuristic behind the constraint in \eqref{eq:propuesta3} is as follows. Because this equation is an upper bound to the maximum singular value associated with the noise, all singular values bigger than $\tau$ correspond to those associated with the signal subspace. Let $s^\star$ be such that $\sigma_{s^\star+1}<\tau<\sigma_{s^\star}$. Then, the model order is at least $s^\star$. Since \eqref{eq:propuesta3} is a loose bound, the signal space may be larger, and some singular values corresponding to the signal subspace may fall under the threshold. Nevertheless, the correct order minimizes \eqref{eq:rhatESTER} or \eqref{eq:rhatSAMOS}
	for $s\ge s^\star$. In other words, we impose a maximum value to the singular values associated with the noise subspace. At the same time we penalize small orders selected with the ESTER rule. When taking $\tau_3$  instead of $\tau_1$ or $\tau_2$, we cannot claim that all singular values bigger than $\tau_3$ correspond to the signal subspace.

\section{Numerical experiments}\label{sec:results}
	 
	We have compared the performance of the model order selection strategy proposed in \eqref{eq:propuesta3} with those rules summarized in section \ref{sec:review}. We have performed Montecarlo simulations for different examples: three were taken from the literature, while the last one is introduced here. %In this section, we list the parameters used in each example, and we describe the experiments performed.  At the end of the section, we discuss the results. 
	
	\subsection{Simulated models}
		Following the usual notation in the literature, we have considered the following model parametrization:
	
		\begin{equation}
			x_k = \sum_{i=1}^{r}a_ie^{\xi_i k} 
			\label{eq:1}
		\end{equation}
		where $\xi_i = 2\pi(\gamma_i+\jmath\nu_i)$. Table \ref{tab:ejemplo} gives the values of the parameters for each example.  Example 1 has two modes located close to each other, and the other two are farther apart. Example 2 is built by adding five more modes to Example 1. In particular, modes $z_4, z_5$, and $z_6$ are clustered in a small region of the complex plane. The modes located in a small region of the complex plane may be confused by the model order selection strategy as a single mode. In Examples 3 and 4 explore further the issue. In both examples, we simulate one single large cluster, which may be obtained when a continuous-time system is digitized using a  high sampling frequency. In Table \ref{tab:ejemplo} we summarize the parameter used in the numerical experimentation. 
	
		\begin{table}[b]
			\begin{tabular}{c|lllll}
				& i & $\nu_i$(rad$^{-1}$) & $\gamma_i$(rad$^{-1}$) & $|a_i|$ & $<a_i$   \\ \hline \hline
				\multirow{4}{.7in}{Example 1  \cite{Andersson2014}} 
				& 1   & -7.68  & -0.274 & 0.4 & -0.93 \\ \cline{2-6}
				& 2   & 39.68 & -0.150 &  1.2 & -1.55 \\ \cline{2-6}
				& 3   & 40.96 &  0.133 &  1.0 & -0.83 \\ \cline{2-6}
				& 4   & 99.84 &  -0.221 & 0.9 & 0.07  \\
				\hline\hline 
				\multirow{7}{.7in}{Example 2  \cite{Andersson2014}} 
				& 1   & -92.16 & 0.177  & 1.0 & 0.42  \\ \cline{2-6}
				& 2   & -7.68  & -0.274 & 1.5 & -0.95 \\ \cline{2-6}
				& 3   & 3.71   & -0.097 & 0.7 & 0.40  \\ \cline{2-6}
				& 4   & 11.90  & -0.116 & 0.6 & 0.02  \\\cline{2-6} 
				& 5   & 14.98  & -0.026 & 1.2 & -1.55  \\\cline{2-6} 
				& 6   & 19.20  & -0.327 & 0.4 & -0.93 \\ \cline{2-6}
				& 7   & 39.68  & -0.150 & 1.0 & -0.83  \\\cline{2-6}
				& 8   & 40.96  &  0.133 & 0.9 &  0.009 \\\cline{2-6}
				& 9   & 99.84  & -0.221 & 0.9 & 0.007   \\
				\hline\hline 
				\multirow{5}{.7in}{Example 3  \cite{Papy2007}} 
				& 1  & 0.2 & -0.01 & 1 & 0.00 \\ \cline{2-6}
				& 2  & 0.3 & -0.02 & 1.0 & 0.00 \\ \cline{2-6}
				& 3   & -0.2 & -0.1 & 2.0 & 0.00 \\ \cline{2-6}
				& 4   & 0.4  & -0.05 & 1.0 & 0.00 \\\cline{2-6}
				& 5   & 0.35 &  0.03 & 1.0 & 0.00  \\
				\hline\hline
				\multirow{6}{.7in}{Example 4} 
				& 1   & -0.22  & -0.01 & 0.97 & -1.78 \\ \cline{2-6}
				& 2   & -0.17 & -0.0037 &  1.58 & 2.89 \\ \cline{2-6}
				& 3   & -0.026 &  -0.0058 &  1.14 & -2.46 \\ \cline{2-6}
				& 4   & 0.0037 &  -0.012 & 0.96 & -1.15  \\
				\cline{2-6}
				& 5   & 0.15 &  -0.0089 & 1.12 & -0.32  \\
				\cline{2-6}
				& 6   & 0.27 &  -0.011 & 1.62 & 0.53  \\
	        	\hline\hline
			\end{tabular}
			\caption{Parameters used in the numerical examples. The  parametrization is as in \eqref{eq:1}.}
			\label{tab:ejemplo}
		\end{table}
		
	\subsection{Analysis of results} 
		For each example in Table \ref{tab:ejemplo}, we compare the following rules:
		\begin{itemize}
			\item $\hat{r}_{\mathrm{thr}}$ where $\tau$ is computed according to \eqref{eq:Gavish_Th};
			\item $\hat{r}_{\mathrm{ESTER}}$ as in \eqref{eq:rhatESTER};
			\item $\hat{r}_{\mathrm{SAMOS}}$ as in \eqref{eq:rhatSAMOS};
			\item $\hat{r}_c$ using the constrained selection rule as in \eqref{eq:propuesta3}  with $\mathcal{J}(s) = J_{SAMOS}(s)$ and $\tau=\tau_1$.	
		\end{itemize}
		Our interest is to accomplish a qualitative comparison for different SNR regimes. Accordingly, we have varied the noise level, and for each SNR, we have considered $N=5000$ independent realizations of \eqref{Eq:noisySignal}. For the analysis, we have calculated the following performance metrics:
		\begin{itemize}
	    	\item Correct order estimation rate: 
	    		\[
	        		\mathrm{COR} = \frac{\text{number of times } \hat{r}=r}{N}
	    		\]
	    	\item Histogram of order estimations. 
		\end{itemize}
	
\begin{figure}[ht!]
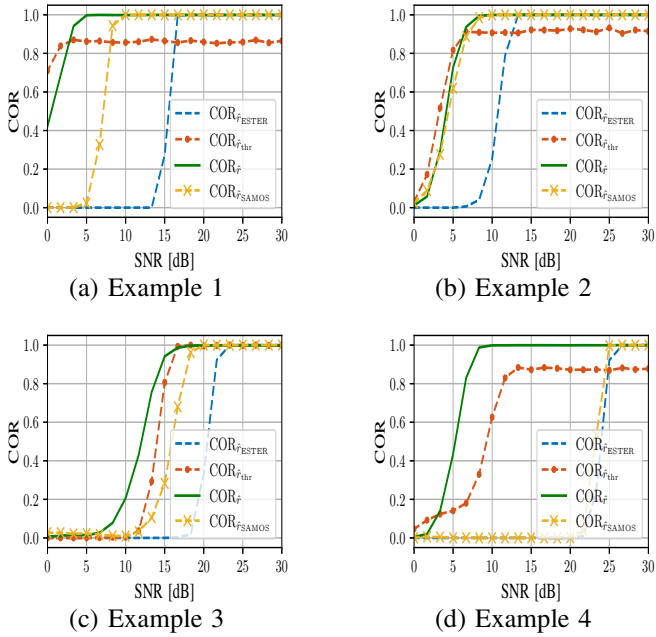

		\begin{minipage}[b]{.45\linewidth}
			\centering
			\centerline{\resizebox{\textwidth}{\linewidth}{\input{img/COR_example_1.pgf}}}
			 \vspace{-0.3cm}
			\centerline{(a) Example 1}\medskip
		\end{minipage}
		\hfill
		\begin{minipage}[b]{.45\linewidth}
			\centering
			\centerline{\resizebox{\textwidth}{\linewidth}{\input{img/COR_example_2.pgf}}}
			\vspace{-0.3cm}
			\centerline{(b) Example 2}\medskip
		\end{minipage}
		\begin{minipage}[b]{0.45\linewidth}
			\centering
			\centerline{\resizebox{\textwidth}{\linewidth}{\input{img/COR_example_3.pgf}}}
			\vspace{-0.3cm}
			\centerline{(c) Example 3}\medskip
		\end{minipage}
		\hfill
		\begin{minipage}[b]{0.45\linewidth}
			\centering
			\centerline{\resizebox{\textwidth}{\linewidth}{\input{img/COR_example_4.pgf}}}
			\vspace{-0.3cm}
			\centerline{(d) Example 4}\medskip
		\end{minipage}
		\caption{Rate of correct order estimation ($\mathrm{COR}$) as a function of the SNR.}
		\label{fig:rates}
	\end{figure}

\begin{figure}[t]
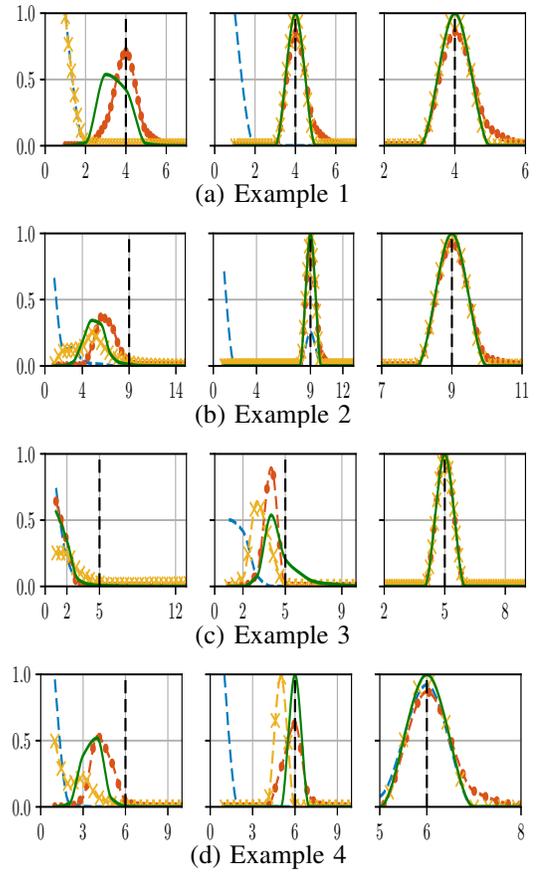

		\centering
		\begin{minipage}[b]{0.8\linewidth}
			\centering
			\centerline{\resizebox{\textwidth}{0.4\linewidth}{\input{img/Histogram_order_ex1.pgf}}}
			\vspace{-0.4cm}
			\centerline{(a) Example 1} %\medskip
		\end{minipage}
		\hfill
		\begin{minipage}[b]{0.8\linewidth}
			\centering
			\centerline{\resizebox{\textwidth}{0.4\linewidth}{\input{img/Histogram_order_ex2.pgf}}}
			\vspace{-0.4cm}
			\centerline{(b) Example 2} %\medskip
		\end{minipage}
		\hfill
		\begin{minipage}[b]{0.8\linewidth}
			\centering
			\centerline{\resizebox{\textwidth}{0.4\linewidth}{\input{img/Histogram_order_ex3.pgf}}}
			\vspace{-0.4cm}
			\centerline{(c) Example 3} %\medskip
		\end{minipage}
		\hfill
		\begin{minipage}[b]{0.8\linewidth}
			\centering
			\centerline{\resizebox{\textwidth}{0.4\linewidth}{\input{img/Histogram_order_ex4.pgf}}}
			\vspace{-0.4cm}
			\centerline{(d) Example 4} %\medskip
		\end{minipage}
		\hfill
		\caption{Histograms of estimated order: $\hat{r}_{ESTER}$ in (- -)blue, $\hat{r}_{c}$ in solid green, $\hat{r}_{thr}$ in (-$\bullet$-) red, and $\hat{r}_{SAMOS}$ in (-x-) yellow. Left columns correspond to SNR = 0dB, center columns to SNR = 10dB, rightmost columns to SNR = 25dB. The vertical dashed black lines denote the actual model order.}	\label{fig:hist}
	\end{figure}

	The results are shown in Figures \ref{fig:rates} and \ref{fig:hist}. In Fig. \ref{fig:hist}, we have performed an interpolation among values of consecutive histogram bins for visualization purposes only.
	By observing Fig.~\ref{fig:rates}, we conclude that both ESTER and SAMOS rules have good performance when the SNR is high. However, they both fail to estimate the actual order when the SNR decreases. Notice  that SAMOS performs  better than ESTER in general. As stated in section \ref{sec:review},  the  objective function in SAMOS is the  average of the sine of all principle angles, while  the objective function in ESTER is only the sine of the maximum principal angle. On the other hand, $\hat{r}_{thr}$ loses performance when the SNR increases. This last observation is explained by  Fig.~\ref{fig:hist} where the histograms show that the optimal threshold rule tends to overestimate the model order when the SNR increases. Notice that the constraint selection rule can maintain the performance of the optimal selection rule at low SNR. Nonetheless, it also recovers the performance of the subspace-based selection rules when the SNR increases. Example 4 is of particular interest to test the behavior of the selection rules when several modes are clustered in a small region. While the constrained selection rule achieves good performance from 5dB on,  ESTER and SAMOS require a SNR higher than 20dB for correct order estimation. Also, the optimal threshold cannot guarantee a correct order selection, even for large SNR.

%	In this case, ESTER and SAMOS require higher than 20 dB for correct order estimation. On the other hand, the optimal threshold cannot guarantee a correct order selection, even for large SNR. 
	
\section{Conclusions}\label{sec:Conclusions}
    We have proposed a new model order selection rule for signals composed of sums of complex exponentials. This scheme benefits from the rotational invariance property of the Hankel matrix, which is beneficial when there is a good separation between the noise and the signal subspaces. But in the low SNR regime, the new scheme can retrieve the actual model order by imposing a maximum bound on the noise contribution to the noisy Hankel matrix. This bound was proposed by analyzing the spectral norm of a random Hankel matrix. To test the performance of the proposed scheme, we have compared it with other selection rules from the literature. The results have shown a better performance of the constrained selection rule for most SNRs in all the examples that were considered.

\appendix

%\section{Proof of lemmas \ref{Lemma:1} and \ref{Lemma:2}}

	\subsection{Proof of lemma \ref{Lemma:1}}\label{Proof:Lemma1}

		Let $\matC_{\w}$ be the  $(m+n-1)\times (m+n-1)$ circulant matrix associated with $\w$. Then, using the definition of  the circulant matrix, we express the $(m\times n)$ Hankel matrix $\H_{\w}$ as %appears in $\matC_{\w}$ in the upper right block with reverse rows. Then we can write

		\[\H_{\w} = \begin{bmatrix} \matT_m & \mathbf{0}_{m\times(n-1)}
		\end{bmatrix}\matC_{\w}\begin{bmatrix} \mathbf{0}_{(m-1)\times n}\\ \matI_n	\end{bmatrix},
		\]
		where $\matT_m$ is an $m\times m$ exchange matrix (backward identity). Clearly,
		\[\begin{aligned}
			\|\H_{\w}\|_2 & \leq \bigg\|\begin{bmatrix} \matT_m & \mathbf{0}_{m\times(n-1)}
			\end{bmatrix}\bigg\|_2\|\matC_{\w}\|_2\bigg\|\begin{bmatrix} \mathbf{0}_{(m-1)\times n}\\ \matI_n
			\end{bmatrix}\bigg\|_2 \\[0.3em]
			& \leq  \|\matC_{\w}\|_2,
		  \end{aligned}
		\] 
where we have used the fact that \[ \big\|\begin{bmatrix} \matT_m & \mathbf{0}_{m\times(n-1)}
			\end{bmatrix}\big\|_2\leq 1, \qquad \bigg\|\begin{bmatrix} \mathbf{0}_{(m-1)\times n}\\ \matI_n
			\end{bmatrix}\bigg\|_2\leq 1, \]
Since $\matC_\w$ is a circulant matrix, its  eigenvalues are $\mathbf{e}_{k+1}^T\matV\w$. Then, 
		\[
		\|\H_{\w}\|_2 \leq \|\matC_{\w}\|_2 = \max_k |\mathbf{e}_{k+1}^T\matV\w|.
		\]
		
%		\[\begin{aligned} 
%			\|\H_{\w}\|_2 & = \max_{\x\in\C^n, \|x\|=1}\|\H_{\w}\x\|_2 \\[0.3em]
%			& = 
%			\max_{\x\in\C^n,\|x\|=1} \bigg\|\begin{bmatrix} \matT_m & \mathbf{0}_{m\times(n-1)}
%			\end{bmatrix}\matC_{\w}\begin{bmatrix} \mathbf{0}_{(m-1)\times n}\\ \matI_n
%			\end{bmatrix}\x\bigg\|_2\\
%								& \leq  \max_{\z\in\C^{m+n-1}, \|z\|=1} \|\matC_{\w}\z\|_2 = \|\matC_{\w}\|_2.
%		  \end{aligned}\] 

	\subsection{Proof of lemma \ref{Lemma:2}}\label{Proof:Lemma2}
	
	 %  \begin{enumerate} 
	        %\item
	        Using lemma \ref{Lemma:1} , we know that $\|\H_\w\|\leq \max_k |\mathbf{e}_{k+1}^T\matV\w|$. Then, 
	        $\P[\|\H_\w\| \leq \tau] \geq \P[\max_k |\mathbf{e}_{k+1}^T\matV\w| \leq \tau]$.
	        
	        Since $\w$ is a complex random vector with gaussian i.i.d. components and $\matV$ is DFT matrix without the normalizing factor $1/\sqrt{m+n-1}$,  the vector $\matV\w$ is an affine transformation which also has gaussian i.i.d.  components. Moreover,  $[\matV\w]_i\sim \mathcal{CN}(0,(m+n-1)\eta^2)$ and  $|[\matV\w]_i|$  is  Rayleigh distributed with parameter $\eta\sqrt{\frac{m+n-1}{2}}$. The distribution of the maximum is
		    \[
		    \begin{aligned}
		    \P(\max_k |\mathbf{e}_{k+1}^T\matV\w|\le \tau) &= \prod_{i=1}^{m+n-1}\P(|[\matV\w]_i|\le \tau) \\ 
		    & = \bigg[1 - e^{-\frac{\tau^2}{(m+n-1)\eta^2}}\bigg]^{m+n-1}
		    \end{aligned}
		    \]
for $\tau \geq 0$.

	       % \item When $\w$ is real vector, its DFT will be even symmetric, i.e. $[\matV\w]_i=[\matV\w]^*_{-i\mod N}$ $\forall i\in\{0,\ldots,N-1\}$. Also, for even $N$, $[\matV\w]_0$ and $[\matV\w]_{N/2}$ are real values, and the DFT will be completely specified by the remainder $N/2-1$ terms. On the other hand, with an odd value of $N$, we have that $[\matV\w]_0$ is real and the DFT will be fully specified with the remaining $(N-1)/2$ terms. Then, it follows that
		
		   % \[[\matV\w]_i \sim \begin{cases} \mathcal{N}(0,N\eta^2) & \text{if $i=0$ or $i = N/2$}\\
		    %     \mathcal{CN}(0,N\eta^2) & \text{otherwise.}
		     %   \end{cases}\] 
		
		    %For $Z_i = |[\matV\w]_i|$, we have that its distribution will be
		
		  %  \[\P[Z_i\leq z] = \begin{cases} \mathrm{erf}\bigg(\frac{z}{\eta\sqrt{2N}}\bigg) & \text{if $i=0$ or $i = N/2$}\\[0.3em] \big[1-e^{-\frac{z^2}{\eta^2N}}\big]\cdot\one\{z\ge 0\} &  \text{otherwise.}
		   % \end{cases} \]
		   % Finally, we get
		
		  %  \[\P(R\le r) = \begin{cases}\mathrm{erf}^2\bigg[\frac{r}{\eta\sqrt{2N}}\bigg]\cdot\bigg[1-e^{-\frac{r^2}{\eta^2N}}\bigg]^{\frac{N}{2}-1} & \text{$N$ even}\\
		   %\mathrm{erf}\bigg[\frac{r}{\eta\sqrt{2N}}\bigg]\cdot\bigg[1-e^{-\frac{r^2}{\eta^2N}}\bigg]^{\frac{N-1}{2}} & \text{$N$ odd.}
		    %    \end{cases}\]
	      
%	   \end{enumerate}
	
\bibliographystyle{IEEEtran}
\bibliography{biblio}

\end{document}